\documentstyle[aps,preprint]{revtex}

\begin{document}
\draft

\title
{\bf Josephson Effect in Superconductors with Odd-Gap Pairing}
\author{M.V.Sadovskii}
\address
{Institute for Electrophysics, Russian Academy of Sciences, \\
Ural Branch, Ekaterinburg 620219, Russia\\
E-mail: sadovski@ief.e-burg.su}
\maketitle

\begin{abstract}
It is shown that the Josephson current in $S-I-S$---geometry is zero, if one or
both superconductors possess the gap odd over $p-p_{F}$.
For the odd-frequency pairing the Josephson current is zero in case of the
tunneling between the traditional and odd-gap superconductor and is finite
for the tunneling between two odd-gap superconductors.
\end{abstract}
\pacs{PACS numbers:  74.20.Fg, 74.50.+r}

\newpage
\narrowtext
Recently a number of non-traditional models of the so called odd-gap pairing
in superconductors were actively studied by a number of authors
\cite{MA,BA,A,KSE,CMT}. These are attractive as possible models for the
interpretation of the physical properties of some existing unusual
superconductors (high-$T_{c}$, heavy-fermion) as well as from the point of
view of the search of the new systems with anomalous properties of
superconducting state. It is most important to formulate some simple
experimental criteria which will unable an unambiguous determination of the
type of corresponding anomalous pairing.

The aim of the present note is to analyze some simple anomalies of
Josephson tunneling in $S-I-S$ structures with odd-gap superconductors.
We shall start with the standard approach to Josephson effect within the
formalism of tunneling Hamiltonian\cite{KY}. This shows that the expression
for the Josephson current through the tunneling contact can be represented in
the following form:
\begin{equation}
I=I_{c}\sin(\phi_{1}-\phi_{2})
\label{1}
\end{equation}
where $\phi_{1}$ and $\phi_{2}$ --- are the phases of superconducting
order-parameter (gap) in $S$-contacts $1$ and $2$, and the critical current
$I_{c}$ is defined by the general expression as:
\begin{equation}
I_{c}=4e\sum_{{\bf p}{\bf q}}|T_{{\bf p}{\bf q}}|^{2}T\sum_{\omega_{n}}
F_{1}^{*}({\bf p},\omega_{n})F_{2}({\bf q},-\omega_{n})
\label{2}
\end{equation}
where $T_{{\bf pq}}$---is the tunneling matrix element for electrons through
the insulating layer, while the anomalous Matsubara-Gorkov functions
$F_{1,2}({\bf p},\omega_{n})$ are determined by:
\begin{equation}
F({\bf p},\omega_{n})=\frac{\Delta({\bf p},\omega_{n})}
{\omega_{n}^2+\xi_{{\bf p}}^2+\Delta^2({\bf p},\omega_{n})}
\label{3}
\end{equation}
Here $\Delta({\bf p},\omega_{n})$---is the gap function,\ $\omega_{n}=(2n+1)
\pi T$ ($T$---temperature), $e$---electronic charge, $\xi({\bf p})=
v_{F}(|{\bf p}|-p_{F})$---electronic excitation energy close to the Fermi level
in the normal state of a superconductor.

Consider first the pairing odd over $p-p_{F}$ \cite{MA,A,KSE}.
In this case we obtain:
\begin{equation}
\Delta({\bf p},\omega_{n})=\Delta(\xi_{\bf p})=-\Delta(-\xi_{\bf p})
\label{4}
\end{equation}
The pairing is isotropic and Eq.(2) is reduced as usual to the following form
\cite{KY}:
\begin{equation}
I_{c}=\frac{T}{\pi eR}\sum_{\omega_{n}}\int_{-\infty}^{\infty}d\xi_{p}
\int_{-\infty}^{\infty}d\xi_{q} F_{1}^{*}(\xi_{p},\omega_{n})F_{2}(\xi_{q},
-\omega_{n})
\label{5}
\end{equation}
where $R$---is the resistance of tunneling contact, expressed via the matrix
element $T_{{\bf pq}}$ averaged over the Fermi surfaces of both
superconductors. It is easy to see that Eq.(5) is actually zero due to the
odd nature of $\Delta(\xi_{p})$ over the variable $\xi_{p}$ (4), both when
only one of superconductors has an odd gap and when both $S$-contacts are
odd. Of course, the transition from Eq.(2) to Eq.(5) and the previous
statement are valid only if we neglect the change of the tunneling matrix
element  $T_{\bf pq}$ on the scale of the order of Fermi energy $E_{F}$
and $I_{c}$ is actually zero neglecting the contribution of the order of
$T^{2}_{pq}/E^{2}_{F}$. From experimental point of view this means almost
total suppression of Josephson tunneling.

Consider now the case odd-frequency pairing\cite{BA,A,CMT}. In this case
\begin{equation}
\Delta({\bf p},\omega_{n})=-\Delta({\bf p},-\omega_{n})
\label{6}
\end{equation}
Then from the general expression (2) it is clear that $I_{c}$ is zero in case
when one of $S$-contacts is the odd-gap superconductor. When both contacts
are odd we obtain the finite expression for $I_{c}$ the explicit form of
which will depend on the specific model of odd-frequency pairing\cite{BA,CMT}.
This may be used to study the properties of the appropriate order-parameter.
The vanishing of Josephson current in the case of contacts of different
parity is connected here with the fact that the odd-frequency gap breaks
the $T$-symmetry, while the tunneling Hamiltonian is obviously invariant
against the time-reversal\cite{KY}.

This work is supported by the State Program of Research on High-Temperature
Superconductivity in the framework of Project $N^{o}$ 93-001.

\end{document}